\documentstyle[12pt]{article}

\input epsf.sty

\begin{document}

\title{Hyperon-rich Matter}

\author{J\"urgen Schaffner\\
The Niels Bohr Institute\\
Blegdamsvej 17\\
DK-2100 Copenhagen}
\maketitle

\begin{abstract}
The phase diagram of nuclear matter offers fascinating features for
heavy ion physics and astrophysics when 
extended to strangeness and antimatter.
We will discuss some regions of this phase diagram
as strange matter at zero temperature, hypermatter at high densities
and the antiworld at high densities and strangeness fraction. 
\end{abstract}

\section{Strange Matter and Hypermatter}

\begin{figure}
\centerline{
\epsfysize=0.35\textheight
\epsfbox{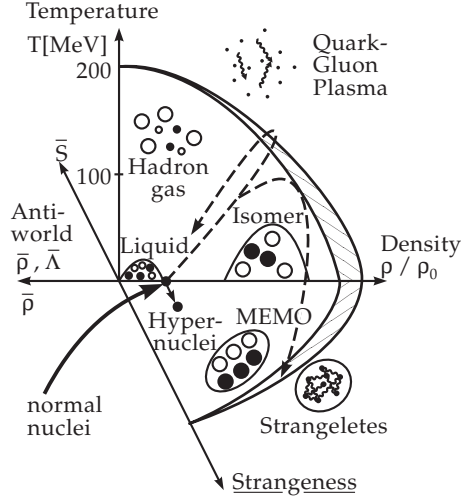}}
\caption{\protect\small
Sketch of the nuclear matter phase diagram with its extensions
to strangeness and to the antiworld.} 
\label{fig:grei1} 
\end{figure}

Relativistic heavy ion collisions provide a promising tool for studying
the physics of strange quark and strange hadronic matter
(see recent review \cite{SM96}). 
Fig.~\ref{fig:grei1} shows schematically the phase diagram of hot, dense
and strange matter. 

Perhaps the only unambiguous way
to detect the transient existence of a quark gluon plasma (QGP)
might be the experimental observation of exotic
remnants, like the formation of strange quark matter (SQM) droplets.
First studies in the context of the MIT-bag model predicted that sufficiently
heavy strangelets might be metastable or even absolutely stable.
The reason for the possible stability
of SQM is related to 
a third flavour degree of freedom, the strangeness.
As the mass of the strange quark is smaller than the Fermi 
energy of the quarks, the total energy of the system is lowered by
adding strange quarks.
According to this picture, 
the number of strange quarks is nearly equal to the number
of massless up or down quarks and saturated SQM is nearly charge neutral.
This simple picture does not hold for small baryon numbers. Finite size effects
shift unavoidably the mass of strangelets to the metastable regime.
Moreover strangelets can have very high charge to mass ratios for
low baryon numbers.
This behaviour is well known from normal nuclei.
Therefore, 
instead of long-lived nearly neutral objects,
strangelet searches in heavy ion experiments 
have to cope with short-lived highly charged objects \cite{Carsten93} !

On the other hand, 
metastable exotic multihypernuclear objects (MEMOs) consisting of
nucleons and hyperons
have been proposed \cite{Sch92}
which extend the periodic system of elements into a new dimension.
MEMOs have remarkably different properties as compared with 
known nuclear matter as
e.g.\ being negatively charged while carrying a positive baryon number!
Even purely hyperonic matter has been predicted \cite{Sch93}.
These rare composites would have a very short lifetime, at the order
of the lifetime of the $\Lambda$.
Central relativistic heavy ion collisions provide a prolific source
of hyperons and hence, possibly, the way for producing MEMOs.
Again, heavy ion experiments looking for these exotic composites
have to deal with very short-lived highly charged objects!

\section{Hyperon-rich Matter in Neutron Stars}

Strangeness, in form of hyperons,
appears in neutron star matter at a moderate density of about
$2-3$ times normal nuclear matter density $\rho_0=0.15$ fm$^{-3}$
as shown by Glendenning within the Relativistic Mean Field (RMF) model
\cite{Glen87}. 
These new species have considerable influences on the equation of state
and the global properties of neutron stars.

On the other side,
much attention has been paid in recent years to the possible onset
of kaon condensation as the other hadronic form of
strangeness in neutron stars.
Most recent calculations based on chiral perturbation theory
\cite{Brown94} show that kaon condensation may set in
at densities of $(3-4)\rho_0$.
Nevertheless, these calculations do not take into account
the presence of hyperons which
may already occupy a large fraction of
matter when the kaons possibly start to condense \cite{Sch94b}. 

Below I present new results from our recent paper \cite{Sch96}, where
the properties of neutron matter with hyperons were studied in detail.
We use the extended version of the 
relativistic mean field (RMF) model and constrain
our parameters to the available hypernuclear data and to the 
kaon nucleon scattering lengths.

\subsection{The Model with Hyperons}

The implementation of hyperons within the RMF approach is straightforward.
SU(6)-symmetry is used for the vector coupling constants
and the scalar coupling constants are fixed to the potential depth of the
corresponding hyperon in normal nuclear matter \cite{Sch93}.
We choose
\begin{equation}
U_\Lambda^{(N)} = U_\Sigma^{(N)} = -30 \mbox{ MeV} \quad , \qquad
U_\Xi^{(N)} = -28 \mbox{ MeV} \quad.
\label{eq:potdep1}
\end{equation}
Note that a recent analysis \cite{Mar95}
comes to the conclusion that the potential changes 
sign in the nuclear interior, i.e.\ being repulsive instead of attractive.
In this case, $\Sigma$ hyperons will not appear at all in our calculations.

The observed strongly attractive $\Lambda\Lambda$ interaction
is introduced by two additional meson fields, the scalar meson $f_0(975)$
and the vector meson $\phi(1020)$.
The vector coupling constants to the $\phi$-field are given by SU(6)-symmetry
and the scalar coupling constants to the
$\sigma^*$-field are fixed by 
\begin{equation}
U^{(\Xi)}_\Xi \approx U^{(\Xi)}_\Lambda \approx
2U^{(\Lambda)}_\Xi \approx 2U^{(\Lambda)}_\Lambda \approx -40 \mbox{ MeV}
\quad .
\label{eq:potdep2}
\end{equation}
Note that the nucleons are not coupled to these new fields.

\subsection{Neutron Stars with Hyperons}

\begin{figure}
\vspace{-0.5cm}
\epsfysize=0.5\textheight
\centerline{\epsfbox{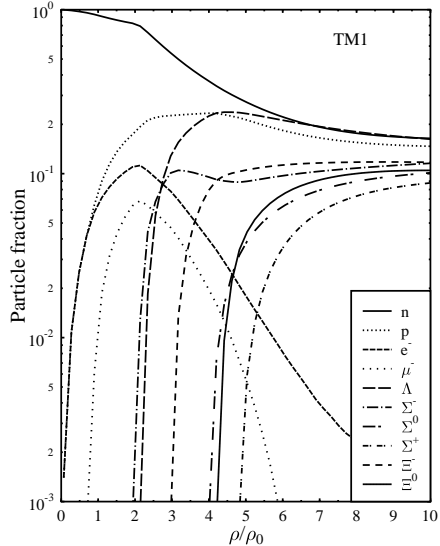}}
\vspace{-1.3cm}
\caption{
The composition of neutron star matter with hyperons which appear
abundantly in the dense interior.
}
\label{fig2}
\end{figure}

Fig.~\ref{fig2} shows the composition of neutron star matter 
for the parameter set TM1 with
hyperons including the hyperon-hyperon interactions.

Up to the maximum density considered here all effective masses
remain positive and no instability occurs. 
The proton fraction
has a plateau at $(2-4)\rho_0$ and exceeds 11\%{} which allows for the
direct URCA process and a rapid cooling of a neutron star.
Hyperons, first $\Lambda$'s and $\Sigma^-$'s, appear at $2\rho_0$, then
$\Xi^-$'s are populated already at $3\rho_0$. The number of electrons
and muons has a maximum here and decreases at higher densities, i.e.\
the electrochemical potential decreases at high densities. 
The fractions of all baryons show a tendency towards
saturation, they asymptotically reach similar values 
corresponding to spin-isospin and hypercharge-saturated matter.
Hence, a neutron star is more likely a giant hypernucleus!

\subsection{Kaon Condensation ?}

In the following
we adopt the meson-exchange picture for the KN-interaction simply
because we use it also for parametrizing the baryon interactions.
We start from the following Lagrangian
\begin{equation}
{\cal L}'_K = D^*_\mu \bar K D^\mu K - m_K^2 \bar K K
- g_{\sigma K} m_K \bar{K}K \sigma
- g_{\sigma^* K} m_K \bar{K}K \sigma^*
\label{eq:modlagr}
\end{equation}
with the covariant derivative
\begin{equation}
D_\mu = \partial_\mu +
ig_{\omega K} V_\mu + ig_{\rho K} \vec{\tau}\vec{R}_\mu + ig_{\phi K}\phi_\mu
\quad .
\end{equation}
The coupling constants to the vector mesons are chosen from
SU(3)-relations.
The scalar coupling constants are fixed by the s-wave KN-scattering lengths.
We have found that this 
leads to an $\bar K$-optical potential around
$U^{\bar K}_{\rm opt} = -(130\div 150)$ MeV at
normal nuclear density for the various parameter sets used.
This is between the two
families of solutions found for Kaonic atoms \cite{Fried93}.
The onset of s-wave kaon condensation is now determined by the condition
$- \mu_e = \mu_{K^-} \equiv \omega_{K^-} (k=0)$.

\begin{figure}
\vspace{-0.5cm}
\epsfysize=0.45\textheight
\centerline{\epsfbox{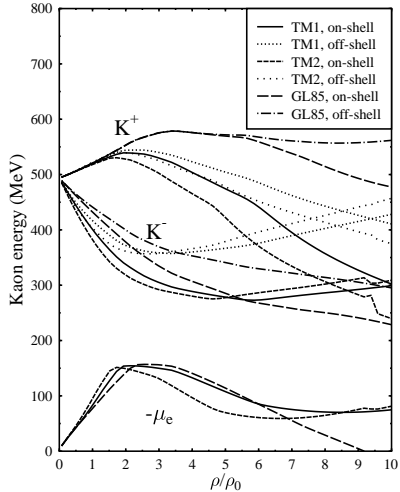}}
\vspace{-1.3cm}
\caption{
The effective energy of the kaon and the antikaon.
and the electrochemical potential.
Kaon condensation does not occur over the whole density region
considered.
}
\label{fig8}
\end{figure}

The density dependence of the K and $\bar K$ effective energies is displayed
in Fig.~\ref{fig8}.
The energy of the kaon is first increasing in accordance with 
the low density theorem.
The energy of the antikaon is decreasing steadily at low densities.
With the appearance of hyperons the situation changes dramatically.
The potential induced by the $\phi$-field cancels the contribution
coming from the $\omega$-meson. Hence, at a certain density the energies of
the kaons and antikaons become equal to the kaon (antikaon) effective mass, 
i.e.\ the curves for kaons and antikaons are crossing at a 
sufficiently high density. At higher densities
the energy of the kaon gets even lower than that of the antikaon!
Since the electrochemical potential never reaches values above 160 MeV here 
antikaon condensation does not occur at all.
We have checked the possibility of antikaon condensation for all parameter
sets and found that at least 100 MeV are missing
for the onset of kaon condensation in contrast to previous 
calculations disregarding hyperons \cite{Brown94}.

\section{Strange Antiworld}

There are evidences for strong scalar and vector 
potentials in nuclear matter. Already in 1956
D\"urr and Teller proposed a relativistic 
model with strong scalar and vector potentials
to explain the saturation of nuclear forces 
\cite{Duerr56} and found a scalar potential of $U_s=a m_N\phi$
where $\phi$ is a scalar field and $a$ is a coupling constant.
This was the first version of the RMF model discussed above, where
$U_s=g_{\sigma} \sigma$, and its extension, 
the Relativistic Br\"uckner-Hartree-Fock (RBHF) \cite{Cel92} calculations,
where the scalar potential is the scalar part of the self-energy of the
nucleon $U_s=\Sigma_s(p_N)$.
In the chiral $\sigma\omega$ model \cite{Bog83} one finds 
$U_s = m_N \sigma /f_\pi - m_N$ which incorporates the (approximate)
chiral symmetry at the underlying QCD Lagrangian.
Here $f_\pi$ is the pion decay constant.
Besides these models based on a hadronic description there exists
effective models dealing with constituent quarks, 
like the Nambu--Jona-Lasinio (NJL) model and models based on QCD sum rules
\cite{Coh91}.
These models can be linked to hadronic observables in the dense medium
by using the low density expansion of the quark condensate
which gives
$U_s=-\frac{m_N \sigma_N}{m_\pi^2 f_\pi^2}\rho_N$, where $\sigma_N\approx 45$ 
MeV is the pion-nucleon sigma term.

Astonishingly, {\em all} these approaches come to the same conclusion,
namely that the scalar potential is as big as
\begin{equation}
U_s = -(350\div 400) \mbox{ MeV } \rho_N/\rho_0
\end{equation}
for moderate densities!
This strong scalar attraction has to be compensated by a strong repulsion
to get the total potential depth of nucleons correct. Hence,
one finds for the vector potential
\begin{equation}
U_v = (300\div 350) \mbox{ MeV } \rho_N/\rho_0
\quad .
\end{equation}
These big potentials are in fact needed to get a correct spin-orbit potential.
The idea of D\"urr and Teller \cite{Duerr56} was that the antinucleons
feel the difference of these two potentials, i.e.
\begin{equation}
U_{\bar N} = U_s - U_v = -(650\div 750) \mbox{ MeV } \rho_N/\rho_0
\end{equation}
which is already comparable to the mass of the nucleon. 
Note that the extrapolation to high densities is quite dangerous as
effects nonlinear in density might get important. It is already known
from RMF models that the scalar potential saturates at high densities
instead of growing steadily.
RBHF calculations show that this might be also true for the vector potential.
With this in mind one can extrapolate to higher densities and finds
that the field potentials get overcritical at $\rho_c=(3-7)\rho_0$
which was first pointed out by Mishustin \cite{Mish90}.
At this critical density the potential felt by the antinucleons is equal
to $U_{\bar N}=2m_N$, the negative energy states are diving in the
positive continuum and this allows for the spontaneous nucleon-antinucleon
pair production. 
This has certain parallels to the spontaneous $e^+e^-$ production 
proposed by Pieper and Greiner \cite{Piep69}.
Assuming SU(6)-symmetry one gets for $\Lambda$'s
\begin{equation}
U^\Lambda_v = (200\div 230) \mbox{ MeV } \rho_N/\rho_0
\end{equation}
and combining with hypernuclear data this gives then
for the total $\bar\Lambda$ potential
\begin{equation}
U_{\bar\Lambda} = U^\Lambda_s - U^\Lambda_v = U_\Lambda - 2U^\Lambda_v 
= -(430\div 500) \mbox{ MeV } \rho_N/\rho_0
\quad .
\end{equation}
In the hyperon-rich medium additional fields will enhance this
potential. Assuming again SU(6)-symmetry one can estimate the vector potential
coming from the $\phi$ meson 
\begin{equation}
V^\Lambda_v = \frac{2}{9} \frac{m_\omega^2}{m_\phi^2} 
U_v \cdot f_s \approx 40 \mbox{ MeV } \rho_B/\rho_0 \cdot f_s
\end{equation}
where $f_s$ is the total strangeness fraction.
The corresponding strange scalar potential is in principle unknown
but definitely higher than the strange vector potential to explain
the strongly attractive $\Lambda\Lambda$ interaction seen in
double $\Lambda$ hypernuclei. Hence one gets at least
an additional $\bar\Lambda$ potential of
\begin{equation}
V_{\bar\Lambda} = V^\Lambda_s - V^\Lambda_v \approx - 120
 \mbox{ MeV } \rho_B/\rho_0 \cdot f_s
\end{equation}
in the hyperon-rich medium.

These strong antibaryon potentials will have certain impacts for heavy ion
reactions. Proposed signals for antiprotons are:
enhanced subthreshold production \cite{Mish90},
change of the slope of the excitation function \cite{Mish90},
apparent higher temperatures \cite{Koch91},
which have indeed been measured at GSI \cite{gsianti}. Nevertheless,
a recent analysis indicates that the antiproton potential might be quite
shallow at normal nuclear density, 
around $U_{\rm\bar p}= -100$ MeV \cite{giessen}.
Possible other signals include:
enhanced antihyperon production \cite{Mish90},
strong antiflow of antibaryons \cite{Jahns},
cold baryons from tunnelling \cite{Mish90},
cold kaons from annihilation in the medium
(the phase space of the reaction 
$\bar\Lambda + {\rm p}\to{\rm K}^+ + \pi's$ is reduced by
$U_{\bar\Lambda} + U_N - U_K \approx -600 \mbox{ MeV } \rho_N/\rho_0$
compared to the vacuum),
enhanced pion production due to the abundant annihilation processes which
would also enhance the entropy.
Definitely, more elaborate work is needed to pin down the possible signals
from the critical phenomenon of the antiworld.

We conclude this section with a brief comment concerning the limitations of the
RMF model. This is clearly an effective model which successfully describes
nuclear phenomenology in the vicinity of the ground state. On the other hand,
this model does not respect chiral symmetry and the quark structure of baryons
and mesons. Also negative energy states of baryons and quantum fluctuations of
meson fields are disregarded. 
These deficiencies may affect significantly the
extrapolations to high temperatures, densities or strangeness contents.

\section*{Acknowledgements}

This paper is dedicated to Prof.\ Walter Greiner
on the occasion of his 60$^{\rm th}$ birthday. 
I am indebted to him for guiding me to the fascinating field
of hypermatter and antimatter and his continuous support.  
I thank my friends and colleagues A. Diener, C.B. Dover, A. Gal, Carsten
Greiner, and especially I.N. Mishustin and H. St\"ocker for their help
and collaboration which made this work possible.


\begin{thebibliography}{99}

\bibitem{SM96}
C. Greiner and J. Schaffner,
in {\em Quark-Gluon Plasma 2}, Ed.\ R.C. Hwa
(World Scientific, Singapore, 1995), p.~635

\bibitem{Carsten93}
C. Greiner, A. Diener, J. Schaffner, H. St\"ocker,
Nucl. Phys. {\bf A566}, 157 (1994)

\bibitem{Sch92}
J. Schaffner, C. Greiner, H. St\"ocker, Phys. Rev. {\bf C46}, 322
(1992)

\bibitem{Sch93}
J. Schaffner, C.B. Dover, A. Gal, C. Greiner, H. St\"ocker,
Phys. Rev. Lett. {\bf 71}, 1328 (1993) and
Ann. of Phys. (N.Y.) {\bf 235}, 35 (1994)

\bibitem{Glen87}
N.K. Glendenning, Astrophys. J. {\bf 293}, 470 (1985)

\bibitem{Brown94}
G.E. Brown, C.-H. Lee, M. Rho, V. Thorsson, Nucl. Phys. {\bf A567}, 937
(1994)

\bibitem{Sch94b}
J. Schaffner, A. Gal, I.N. Mishustin, H. St\"ocker, W. Greiner,
Phys. Lett. {\bf B334}, 268 (1994)

\bibitem{Sch96}
J. Schaffner and I.N. Mishustin, 
Phys. Rev. {\bf C53}, 1416 (1996)

\bibitem{Mar95}
J. Mares, E. Friedman, A. Gal, B.K. Jennings,
Nucl. Phys. {\bf A594}, 311 (1995)

\bibitem{Fried93}
E. Friedman, A. Gal, C.J. Batty, Phys. Lett. {\bf B308}, 6 (1993);
Nucl. Phys. {\bf A579}, 518 (1994)

\bibitem{Duerr56}
H.P. D\"urr and E. Teller, Phys. Rev. {\bf 101}, 494 (1956)

\bibitem{Cel92}
L.S. Celenza, A. Pantziris, C.M. Shakin, W.D. Sun, 
Phys. Rev. {\bf C45}, 2015 (1992)

\bibitem{Bog83}
J. Boguta, Phys. Lett. {\bf 120B} 34 (1983)

\bibitem{Coh91}
T.D. Cohen, R.J. Furnstahl, D.K. Griegel, 
Phys. Rev. Lett. {\bf 67}, 961 (1991)

\bibitem{Mish90}
I.N. Mishustin, Yad. Fiz. (Sov. J. Nucl. Phys.) {\bf 52}, 1135 (1990),
J. Schaffner, I.N. Mishustin, L.M. Satarov, H. St\"ocker, W. Greiner,
Z. Phys. {\bf A341}, 47 (1991),
I.N. Mishustin, L.M. Satarov, J. Schaffner, H. St\"ocker, W. Greiner,
J. Phys. {\bf G19}, 1303 (1993)

\bibitem{Piep69}
W. Pieper and W. Greiner, Z. Phys. {\bf A218} 327 (1969)

\bibitem{Koch91}
V. Koch, G.E. Brown, C.M. Ko,
Phys. Lett. {\bf B265}, 29 (1991)

\bibitem{gsianti}
A. Schr\"oter, E. Berdermann, H. Geissel, P. Kienle, W. K\"onig, A. Gillitzer,
J. Homolka, F. Schumacher, H. Str\"oher, B. Povh,
Nucl. Phys. {\bf A553} (1993) 775c

\bibitem{giessen}
St. Teis, W. Cassing, T. Maruyama, U. Mosel,
Phys. Rev. {\bf C50}, 388 (1994)

\bibitem{Jahns}
A. Jahns, C. Spieles, H. Sorge, H. St\"ocker, W. Greiner,
Phys. Rev. Lett. {\bf 72}, 3464 (1994)

\end{thebibliography}
\end{document}